\documentclass{aa}
\usepackage[T1]{fontenc}
\usepackage{eepic,epsfig}
\usepackage{graphicx}
\usepackage{amssymb}
\usepackage{float}
\usepackage{color}
\usepackage{geometry}
\usepackage{marginnote}
\usepackage{amsmath}
\usepackage{natbib}
\usepackage{bibentry}
\usepackage{array} 
\usepackage[varg]{txfonts}
\usepackage{caption}
\usepackage{bm}
\usepackage{subcaption}
\usepackage{xcolor}
\usepackage[super]{nth}
\usepackage[labelfont=color]{caption}
\DeclareMathSymbol{\varOmega}{\mathord}{letters}{"0A}
\DeclareMathSymbol{\varPsi}{\mathord}{letters}{"09}
\DeclareMathSymbol{\varPhi}{\mathord}{letters}{"08}
\DeclareMathSymbol{\varGamma}{\mathord}{letters}{"00}
\DeclareMathSymbol{\varPi}{\mathord}{letters}{"05}
\DeclareMathSymbol{\varLambda}{\mathord}{letters}{"03}
\DeclareCaptionFont{red}{\color{red}}
\usepackage{mathrsfs}
\begin{document}
\title{Erosion of planetesimals by gas flow}
\author{Noemi Schaffer \inst{1}
\and Anders Johansen \inst{1}
\and Lukas Cedenblad \inst{2}
\and Bernhard Mehling \inst{3}
\and Dhrubaditya Mitra \inst{2}}
\institute{Lund Observatory, Department of Astronomy and Theoretical Physics, Lund University, Box 43, 22100 Lund, Sweden \and NORDITA, Stockholm,
Roslagstullsbacken 23
106 91 Stockholm, Sweden \and Department of Physics , Gothenburg University, SE-41296 Gothenburg, Sweden \\ \email{noemi.schaffer@astro.lu.se}}
\date{}
\abstract{The first stages of planet formation take place in protoplanetary disks that are largely made up of gas. Understanding how the gas affects planetesimals in the protoplanetary disk is therefore essential. In this paper, we discuss whether or not gas flow can erode planetesimals. We estimate how much shear stress is exerted onto the planetesimal surface by the gas as a function of disk and planetesimal properties. To determine whether erosion can take place, we compare this with previous measurements of the critical stress that a pebble-pile planetesimal can withstand before erosion begins. If erosion takes place, we estimate the erosion time of the affected planetesimals. We also illustrate our estimates with two-dimensional numerical simulations of flows around planetesimals using the lattice Boltzmann method. We find that the wall shear stress can overcome the critical stress of planetesimals in an eccentric orbit within the innermost regions of the disk. The high eccentricities needed to reach erosive stresses could be the result of shepherding by migrating planets. We also find that if a planetesimal erodes, it does so on short timescales. For planetesimals residing outside of $1 \ \rm{au}$, we find that they are mainly safe from erosion, even in the case of highly eccentric orbits.}
\titlerunning{Erosion of planetesimals by gas flow}
\maketitle 

\section{Introduction}

Planet formation begins with the collisional growth of micrometer sized dust into millimeter-centimeter sized pebbles. Through processes such as the streaming instability pebbles then can be concentrated into planetesimals with characteristic sizes of approximately $100 \ \rm{km}$ \citep{Youdin2005, Youdin2007, Johansen2007, Simon2016}. Smaller planetesimals may form as a result of fragmentation of such large planetesimals, boosting core formation by planetesimal accretion \citep{Dangelo2014}. Other destructive interactions can take place in planetesimal-planetesimal collisions as well as in collisions between a planetesimal and remaining pebbles - the outcome of such collisions among others may be fragmentation \citep{Wilkinson2008} and erosion \citep{Guttler2010,Guilera2014,Syed2017,Jansson2017}. Since the initial stages of the planet formation process take place while the gas is still present in the disk, it is essential to also understand whether the interaction between the gas and planetesimals can also cause erosion. 

Apart from a few previous works, the effect of gas on the erosion of disk solids has mostly been ignored. \cite{Paraskov2006} performed experiments and numerical calculations on the erosion of dust aggregates and found that solids on eccentric orbits are likely eroded in typical protoplanetary disk settings. Through parabolic flight experiments on the wind induced erosion of a dust bed, \cite{Demirci2019} also show that within the inner protoplanetary disk the wall stress from the gas is enough to erode planetesimals.

In this paper, we study where in protoplanetary disks erosion due to gas flow is significant. We use an analytical model based on boundary-layer theory. We find the combination of the gas disk properties and planetesimal sizes that result in erosion in our model. For the planetesimal sizes where erosion is likely to be significant, we estimate the amount of time it takes to erode a given solid. We also build numerical models of the solid-gas interaction using the lattice Boltzmann method that provides us with profiles of wall stress along the solid surface as a function of the Reynolds number. 

In Sect. \ref{boundary_layer} we cover the theoretical aspect of the problem. In particular, we calculate the stress exerted by the gas onto planetesimals with varying sizes in typical protoplanetary disk settings. In Sect. \ref{numeric_model} we present the results of the lattice Boltzmann method to perform numerical simulations of the gas flow around a solid obstacle. We summarize our result in Sect. \ref{summary} and discuss the lattice Boltzmann method in detail in Appendix \ref{Appendix1}.

\section{Boundary layer model}
\label{boundary_layer}

In order to understand how protoplanetary disk solids erode due to gas flow, we calculate the wall shear stress, $\tau_{\rm{w}}$, that is exerted by the gas onto the planetesimal surface. This stress is the tangential force per unit area exerted onto the planetesimal. Following \cite{Jager2017}, we assume that if the wall shear stress exceeds a threshold value, $\tau_{\rm{cr}}$, the solid starts to erode. The rate of mass loss per unit area, $\mathscr{F}$, which we also call the rate of erosion, is proportional to $\tau_{\rm{w}} - \tau_{\rm{cr}}$, i.e.,

\begin{align}
\mathscr{F}=
\left\{
\begin{array}{ll}
      \kappa_{\rm{er}}(\tau_{\rm{w}}-\tau_{\rm{cr}}) & \tau_{\rm{w}} > \tau_{\rm{cr}} \\
      0 & \tau_{\rm{w}} \leq \tau_{\rm{cr}}. \\
\end{array} 
\right.
\label{massloss}
\end{align}

\noindent In Eq. \eqref{massloss} $\kappa_{\rm{er}}$ is a constant of proportionality which we call the erosion coefficient. Both the erosion coefficient and the critical stress depend on the material property of the solid, e.g., its composition and porosity. Equation \eqref{massloss} is purely empirical. This and similar erosion equations have been used for a long time - since the pioneering work by \cite{Shields1936} - in studies of erosion of river beds (see e.g., \cite{Subhasish2014} for an introduction). To find an estimate of the threshold stress can be quite a difficult task. For river beds a commonly used assumption is that the threshold stress is set by gravity and friction. For small planetesimals gravity is expected to play almost no role. The threshold stress should be set by surface forces.

For the erosion coefficient we use measurements from the experiments performed by \cite{Demirci2019}. They studied the wind erosion of a bed of spherical glass beads in a setting that resembles protoplanetary disks moderately well. Near the erosion threshold, they measured a mass loss rate of about $0.6 \ \rm{kg \ m}^{-2} \ \rm{s}^{-1}$. This and the critical shear stress value set by the data from \cite{Demirci2019} (sum of gravity and cohesion) results in $\kappa_{\rm{er}}$$\sim$$150 \ \rm{s/m}$, given that $\tau_{\rm{w}}-\tau_{\rm{cr}} \sim 4 \times 10^{-4}$ at the threshold friction velocity. 
As seen in Eq. \eqref{massloss}, the magnitude of $\kappa_{\rm{er}}$ is responsible for setting the efficiency of the rate of mass change. Thus, once experiments with conditions that resemble true protoplanetary disk settings are performed, $\mathscr{F}$ should be adjusted accordingly.

The wall shear stress exerted by the flow on the solid surface is defined as

\begin{equation}
\tau_{\rm{w}}=\tau(y=0)=\mu\frac{\partial u}{\partial y} \Bigg \vert_{y=0},
\label{wallshearstress}
\end{equation}

\noindent where $\mu = \rho \nu$ is the dynamic viscosity, with $\rho$ denoting the gas density and $\nu$ the kinematic viscosity. We choose our $y$ axis along the wall normal direction and $u$ is the streamwise component of the velocity. We approximate Eq. \eqref{wallshearstress} as 

\begin{equation}
\tau_{\rm{w}} = \mu \frac{u}{\delta},
\label{wallshearstress_simple}
\end{equation}

\noindent where $\delta$ is the thickness of the boundary layer around a solid with size $L$ and $U$ is the typical velocity of the mean flow. If the boundary layer is laminar (see e.g., \cite{Landau1987}, chapter IV), its thickness can be estimated as 

\begin{equation}
\delta=\sqrt{\frac{\nu L}{U}}.
\label{delta}
\end{equation} 

\noindent In terms of protoplanetary disk parameters the wall shear stress defined in Eq. \eqref{wallshearstress_simple} becomes

\begin{equation}
\tau_{\rm{w}}=\rho \nu^{1/2} U^{3/2} L^{-1/2},
\label{shear_PPD}
\end{equation}

\noindent where the density and the kinematic viscosity of the disk are set according to the adopted protoplanetary disk model (e.g. the minimum mass solar nebula, \cite{Hayashi1981}). The parameter, $U$, in this case is defined as the speed difference between the gas and the planetesimal, i.e., the headwind. For a planetesimal on an eccentric orbit, the headwind is not a constant but depends on the details of the planetesimal's orbit. For a planetesimal, in a orbit with zero inclination at perihelion, the headwind is given by \citep{Adachi1976}

\begin{equation}
U=v_{\rm{k}}\sqrt{e^2+\eta^2},
\label{speed}
\end{equation}

\noindent where $\eta$ is the proportional to the ratio of the squares of the mean thermal speed of the gas molecules and the Keplerian speed \citep{Adachi1976}. The value of $\eta$ is estimated to be approximately $10^{-3}$. In Eq. \eqref{speed}, $v_{\rm{K}}=\sqrt{\frac{GM_{\odot}}{r}}$ is the Keplerian speed at a distance, $r$, from the central star with solar mass, $M_{\odot}$. 

The critical shear stress that a protoplanetary-disk solid can withstand is determined by the strength of its surface layer. \cite{Skorov2012} model the tensile strength of ice-free outer layers of comet nuclei given that they are agglomerates of dust aggregates. They conclude that the effective tensile strength of such a layer is 

\begin{equation}
T_{\rm{eff}} = 1.6 \ \rm{Pa} \ \phi_{\rm{p}} \Bigg(\frac{\mathit{a}}{1 \ \rm{mm}} \Bigg)^{-2/3},
\end{equation}

\noindent where $\phi_{\rm{p}}=1-P$ is the volume filling factor and $a$ is the size of the aggregate. \cite{Baer2011} present the bulk porosity, $P$, of more than 50 main-belt asteroids as $P \sim 0.1-0.7$. Taking $P=0.4$ as the average porosity, the effective tensile strength of a $1 \ \rm{mm}$ dust aggregate results in $\tau_{\rm{cr}} = 0.96 \ \rm{Pa}$. We also compare the wall shear stress exerted by the gas to the critical shear stress determined by experiments presented in \cite{Paraskov2006}. They performed experiments on centimeter sized dust aggregates in a wind tunnel, where they studied the effect of gas flow on the aggregates, which consisted of about $500 \ \mu \rm{m}$ dust granules. They found that in the case of a planetesimal on an eccentric orbit $\tau_{\rm{cr}}=25 \ \rm{Pa}$.

\begin{figure}[!t]
\centering
\includegraphics[width=1\columnwidth]{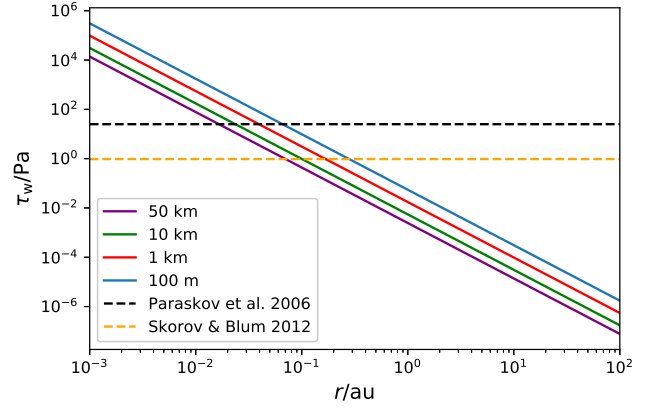}\caption{The scaling of the the wall shear stress exerted on planetesimals with varying sizes as a function of orbital distance, $r$. Here, we assume an orbital eccentricity of $e=0.1$. The black dashed line marks the $25  \ \rm{Pa} $ limit for erosion given by \cite{Paraskov2006}. The orange dashed line marks the $0.96 \ \rm{Pa}$ limit of the tensile strength of dust layers on comets given by \cite{Skorov2012}. The solid curves with varying colors correspond to varying planetesimal sizes. }\label{shear_allsize}
\end{figure}

Figure \ref{shear_allsize} shows the comparison between the wall shear stress from our boundary layer model and the critical shear stress from \cite{Paraskov2006} (marked with black dashed line) and \cite{Skorov2012} (marked with orange dashed line) as a function of orbital distance. The eccentricity of the planetesimal orbit is taken as $e=0.1$ in a minimum mass solar nebula model. The four solid curves in Fig. \ref{shear_allsize} correspond to different planetesimal sizes. Overall, $\tau_{\rm{w}}$ increases with decreasing planetesimal size and with decreasing distance from the central star. Thus, erosion due to gas flow is likely to be most efficient in the inner protoplanetary disk.

\begin{figure*}[!t]
   \centering
   \begin{subfigure}[b]{0.325\textwidth}
   \includegraphics[width=1\columnwidth]{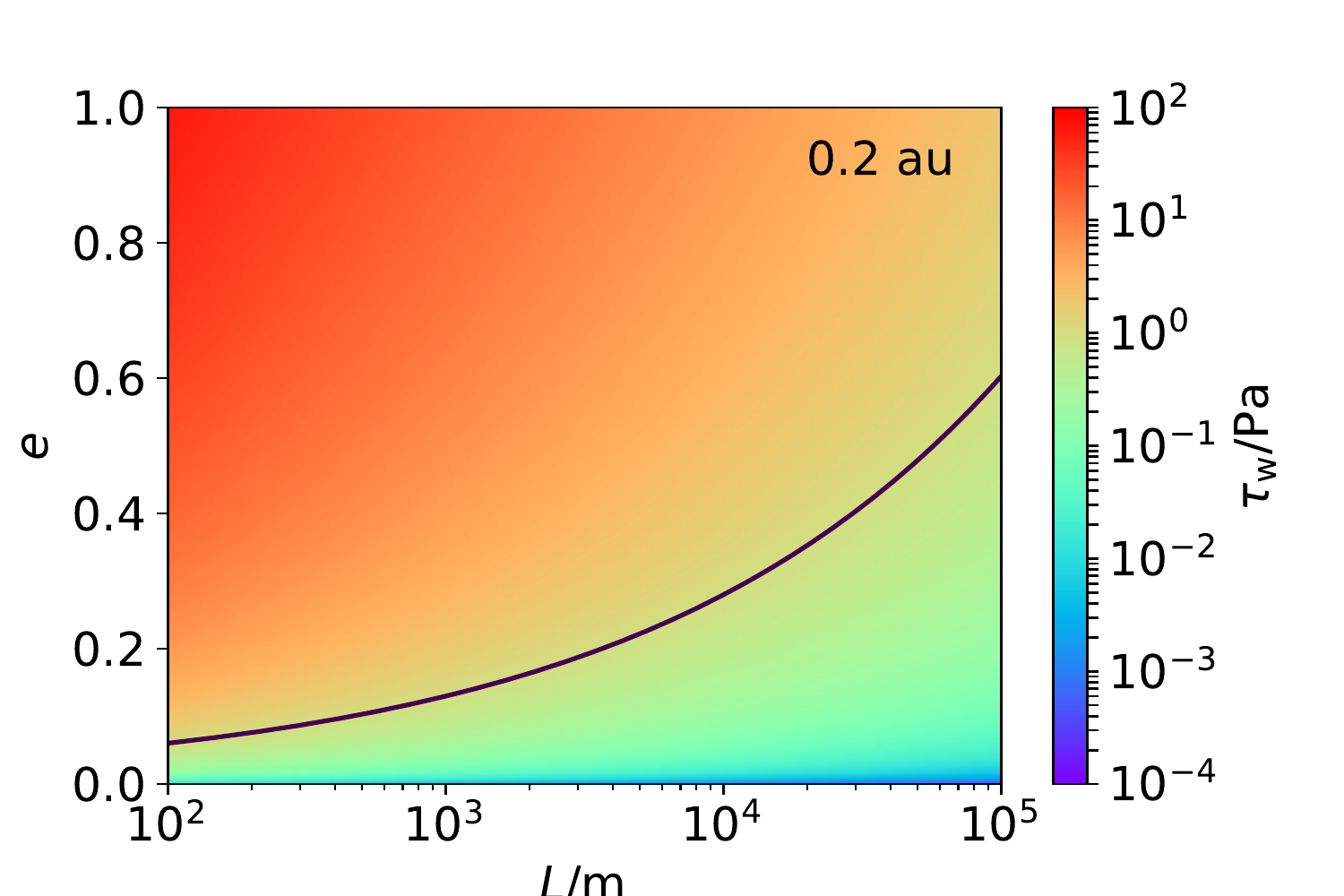}\caption{}
   \label{L_e_02}
   \end{subfigure}
   \begin{subfigure}[b]{0.325\textwidth}
   \centering
   \includegraphics[width=1\columnwidth]{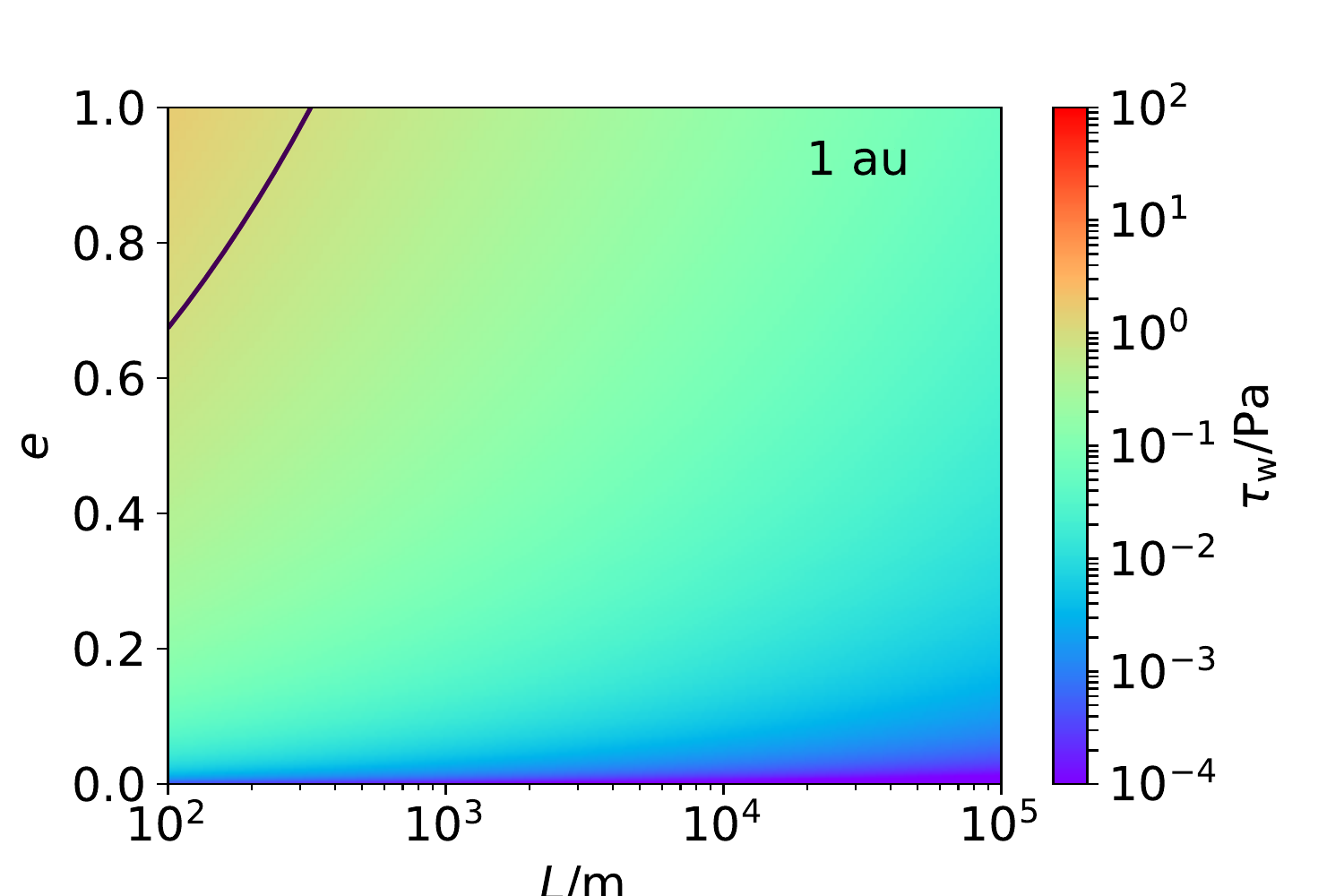}\caption{}
   \label{L_e_1}
   \end{subfigure}
   \begin{subfigure}[b]{0.325\textwidth}
  \centering   
  \includegraphics[width=1\columnwidth]{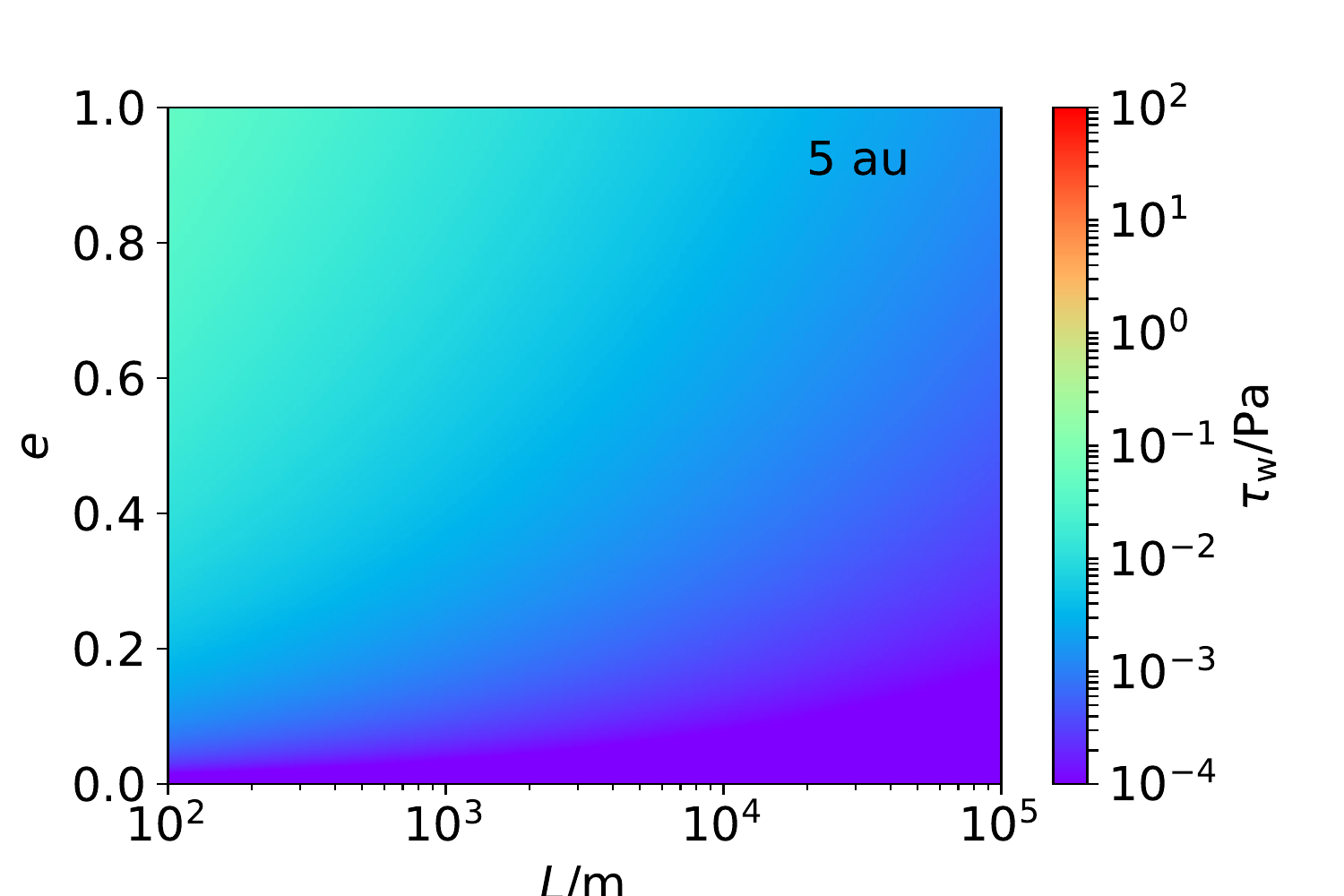}\caption{}
   \label{L_e_5}
   \end{subfigure}
\caption{Wall shear stress as a function of planetesimal size, $L$, and orbital eccentrity, $e$, at an orbital distance of (a) $r=0.2 \ \rm{au}$, (b) $r=1 \ \rm{au}$ and (c) $r=5 \ \rm{au}$. The black solid curve corresponds to $\tau_{\rm{w}}=0.96 \ \rm{Pa}$, which we take as the critical shear stress  that a planetesimal can withstand before it begins to erode \citep{Skorov2012}.}
\label{L_e}
\end{figure*}

The wall shear stress as defined in Eq. \eqref{shear_PPD} is a function of the disk density and viscosity and the Keplerian velocity, which are all functions of the radial coordinate of the planetesimal, $r$. The velocity in addition also depends on the eccentricity of the orbit as stated in Eq. \eqref{speed}. Finally, $\tau_{\rm{w}}$ depends on the size of the planetesimal itself. Based on Eqs. \eqref{shear_PPD} and \eqref{speed} we expect that erosion due to the wall shear stress exerted by the gas flow increases with orbital eccentricity and is most effective in the inner disk. To understand the relationship between the wall shear stress and $r$, $e$ and $L$ we can turn to Fig. \ref{L_e}, where we show our model predictions for the wall shear stress as a function of orbital eccentricity and planetesimal size at $r=0.2 \ \rm{au}$, $r=1 \ \rm{au}$ and $r=5 \ \rm{au}$. The black curves correspond to $\tau_{\rm{w}} = 0.96 \ \rm{Pa}$, which is the critical shear stress from \cite{Skorov2012}. According to Fig. \ref{L_e_02}, at $r=0.2 \ \rm{au}$ most planetesimal sizes seem to experience erosion. Once the realistic eccentricity of the planetesimal with the given size is taken into account as well, we can conclude that large planetesimals ($\sim$$100 \ \rm{km}$) are only likely to experience erosion in the inner regions of the disk, given the presence of a large planet to excite the eccentricities. Planets tend to migrate as they interact with the gas disk \citep{Walsh2012,Morbidelli2012,Batygin2015,Ida2018,Johansen2019}. As they migrate, they can shepherd planetesimals in the inner disk and excite their eccentricities \citep{Tanaka1999}. Smaller planetesimals are likely eroded by the gas flow as well as shown in Fig. \ref{L_e_02}. Figure \ref{L_e_1} shows that beyond $r=1 \ \rm{au}$ planetesimals are not expected to be eroded, since the wall shear stress only exceeds $\tau_{\rm{cr}}$ where the planetesimals are likely too small to acquire high enough eccentricities in realistic protoplanetary disk settings. At a distance of already $5 \ \rm{au}$, it is highly unlikely that any planetesimal would be eroded by gas flow, as shown in Fig. \ref{L_e_5}. Consequently, the ideal regime for erosion due to gas flow is in the inner $r \approx 1 \, \rm{au}$ of protoplanetary disks. Erosion due to gas flow is most significant for small planetesimals of $\sim 100 \, \rm{m}$ and large planetesimals of $\sim 100 \, \rm{km}$ on eccentric orbits.

\begin{figure}[!t]
  \centering
  \includegraphics[width=1\columnwidth]{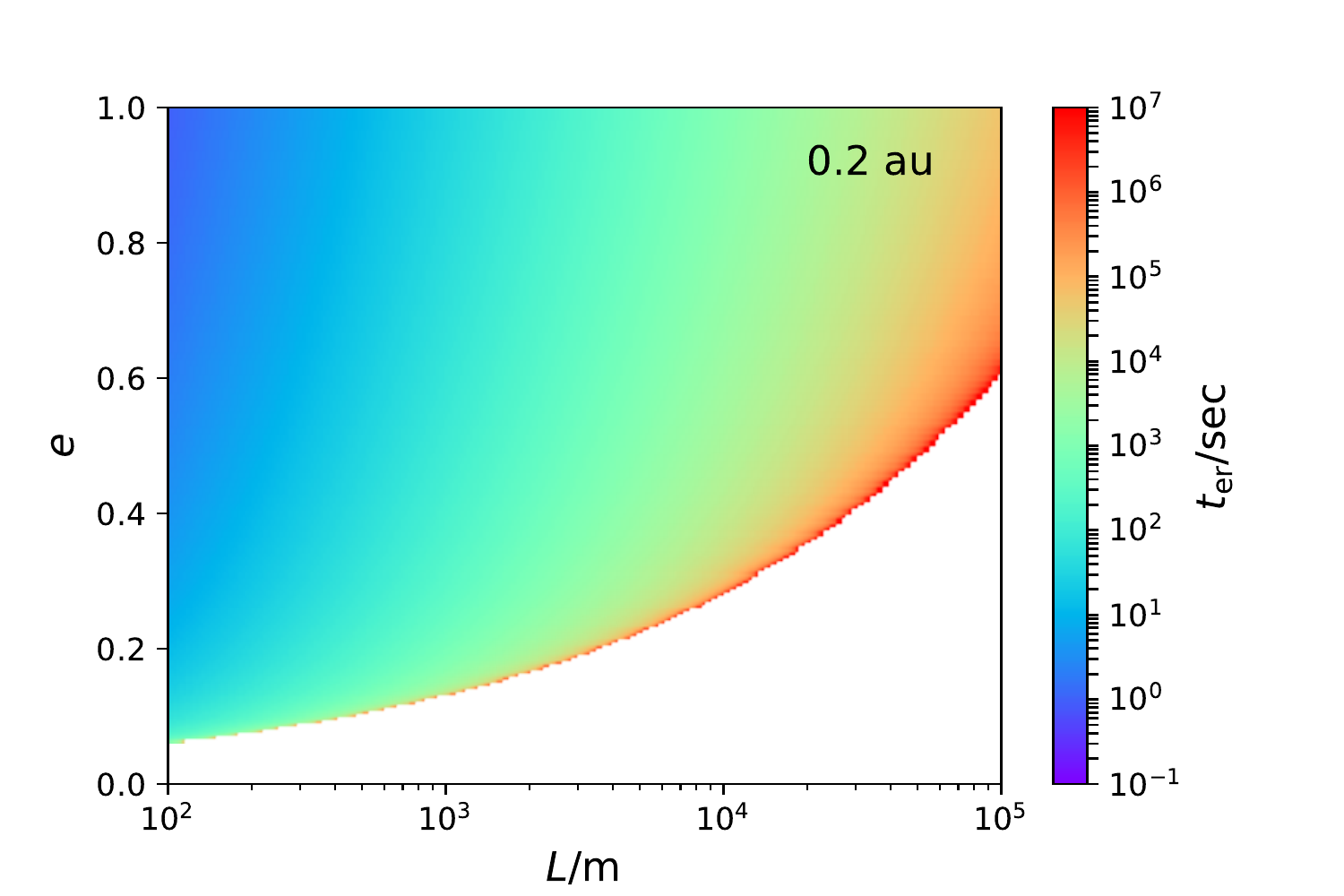}\caption{Erosion time as a function of planetesimal size, $L$, and orbital eccentrity, $e$, at an orbital distance of $r=0.2 \ \rm{au}$. The erosion time is defined only, where $\tau_{\rm{w}} > \tau_{\rm{cr}}$.}\label{t_er_02}
\end{figure}

It is non-trivial to estimate the time it takes for a planetesimal to erode away because as the planetesimal erodes, its size changes and so does the flow around it. As a result, the wall shear stress changes as well. Here, we present only an estimate that is based on the wall stress calculated from the initial size of the planetesimal such that the erosion time is

\begin{equation}
t_{\rm{er}} = \frac{m}{\mathscr{F}A},
\label{erosion_time}
\end{equation}

\noindent where the total planetesimal mass is 

\begin{equation}
m=\rho_{\bullet} V,
\end{equation}

\noindent where $\rho_{\bullet}$ is the material density of the solid and the volume of the planetesimal is $V=\frac{4}{3} \pi \Big (\frac{L}{2} \Big )^3$. The area of the planetesimal in Eq. \eqref{erosion_time} is $A=4 \pi \Big(\frac{L}{2}\Big)^2$. We take $\rho_{\bullet} = 5.4 \times 10^2 \ \rm{kg/m^3}$, which is the approximate bulk density of 67P/Churyumov–Gerasimenko \citep{Patzold2019}. The erosion time is defined only in regions, where $\tau_{\rm{w}} > \tau_{\rm{cr}}$. Given that a planetesimal is within the regime where erosion due to gas flow is possible, the erosion is very fast, as shown in Fig. \ref{t_er_02}. The figure also shows that $t_{\rm{er}}$ is a strong function of planetesimal size, such that it decreases with decreasing $L$. 

As shown in Fig. \ref{t_er_02}, the erosion happens on a short timescale, whereas the radial drift timescale is on the order of $\sim$$10^2-10^4 \ \rm{years}$ in the inner disk \citep{Youdin2010}. As a result, if the planetesimals come in gravitational contact with a migrating planet and thus acquire excited eccentricities, they erode  away almost instantaneously, without experiencing radial drift.

\section{Numerical model}
\label{numeric_model}

To model the stress on the planetesimal we have developed a new code using the lattice Boltzmann method, which is described in Appendix \ref{Appendix1} in detail. Usually, most codes used in astrophysics do not allow us to solve for flows with arbitrary boundaries, which is necessary in this problem. The immersed boundary method has been used in \cite{Mitra2013} to calculate flows around spherically shaped planetesimals facing a headwind. However, during erosion the shape of the body changes dynamically, thus we use the lattice Boltzmann method. Here, we do not implement erosion in our code and do not solve the flow for a dynamically changing boundary, but calculate the wall shear stress for just the initial shape and size of the planetesimal. We still use the lattice Boltzmann method, since it has the possibility of extending our calculations to the dynamically changing case.

We build a two-dimensional code to simulate the flow around a planetesimal and measure the wall shear stress exerted onto its surface. In order to understand how the flow properties influence the outcome, we then compare the results as a function of the Reynolds number. This is a non-dimensional number and is a measure of the balance between inertial and viscous forces defined as 

\begin{equation}
Re=\frac{UL}{\nu}.
\label{Reynolds}
\end{equation}

\noindent Through the Reynolds number of the numerical model (see Eq. \eqref{Reynolds_code}), one can scale the numerical setup to physical protoplanetary disk values. The Reynolds number can be written as 

\begin{equation}
Re \sim \frac{ev_{\rm{K}} L}{c_{\rm{s}} \lambda} \sim \frac{e \varOmega r L}{c_{\rm{s}}} n_{\rm{H}_2} \sigma_{\rm{H}_2} \sim e \varOmega^2 r L \frac{\varSigma}{\sqrt{2 \pi} c^2_{\rm{s}}} \frac{\sigma_{\rm{H}_2}}{m_{\rm{H}_2}},
\label{Re_sound}
\end{equation}

\noindent where $\lambda$ is the mean free path of the gas, $\varOmega$ is the Keplerian angular frequency and $\mathbf{\varSigma= 1700 \, \rm{ g cm}^{-2}\,(\mathit{r}/{\rm AU})^{-1.5}}$ is the surface density of the disk. In Eq. \eqref{Re_sound}, the sound speed is denoted with $c_{\rm{s}}$ and its value is set according to the MMSN model at a distance of $r=1 \rm{au}$. The number density of hydrogen molecules is $n_{\rm{H}_2}=\rho_{\rm{H}_2}/m_{\rm{H}_2}$. For the collisional cross section of the hydrogen molecule, we use $\sigma_{\rm{H}_2}=2 \times 10^{-15} \, \rm{cm}^{2}$ from \cite{Chapman1970}. The mass of the molecule is $m_{\rm{H}_2}$. In the minimum mass solar nebula $Re$ thus scales as 

\begin{equation}
Re \sim 5 \times 10^5 \Bigg(\frac{e}{0.1}\Bigg) \Bigg(\frac{L}{10^3 \ \rm{m}} \Bigg) \Bigg(\frac{M}{M_{\odot}} \Bigg) \Bigg(\frac{r}{\rm{au}} \Bigg)^{-3}.
\label{Re_pp}
\end{equation}

\begin{figure}[!t]
  \centering
  \includegraphics[width=1\columnwidth]{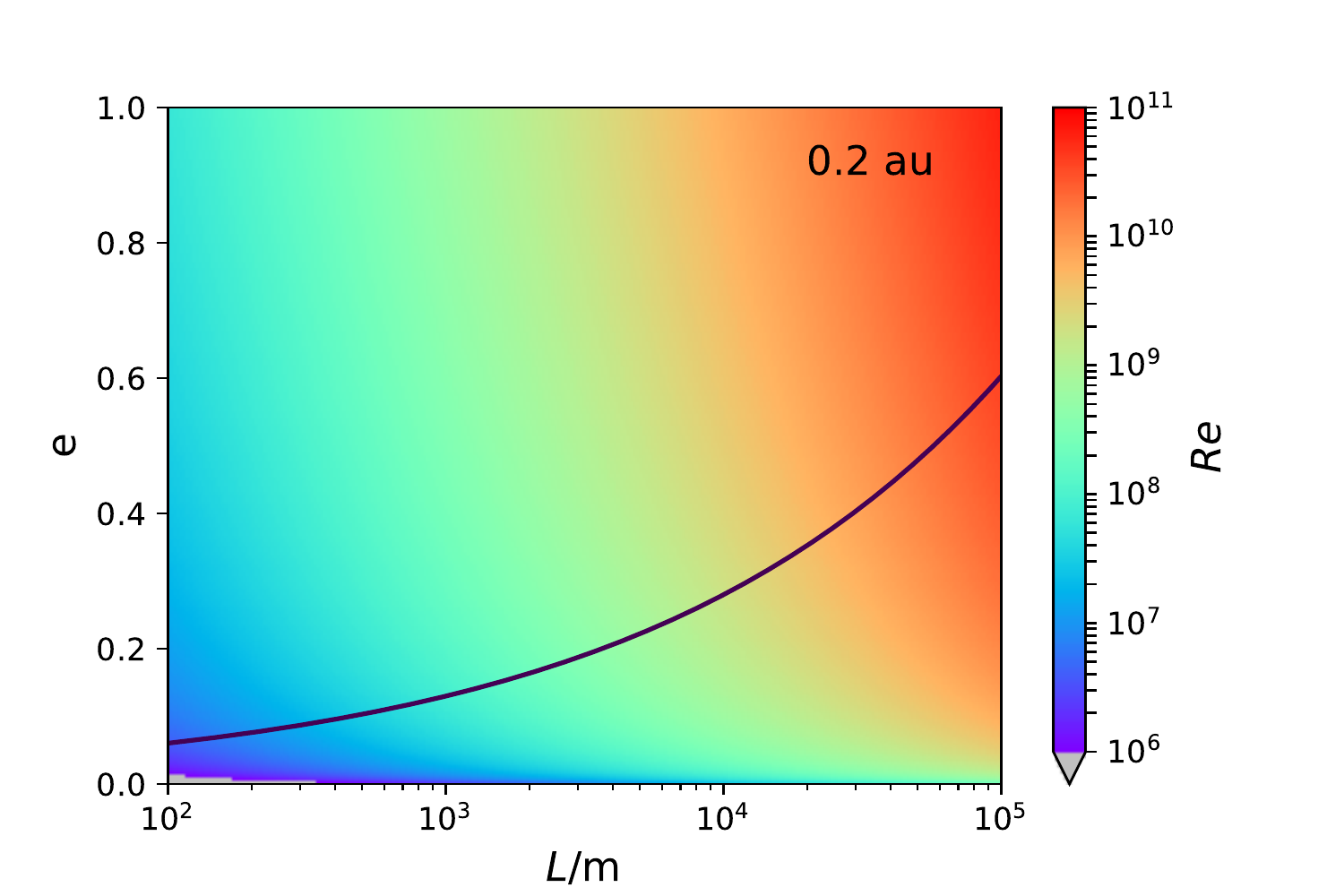}\caption{Reynolds number as a function of planetesimal size and eccentricity at an orbital distance of $r = 0.2 \ \rm{au}$ accoding to the minimum mass solar nebula model. The black curve corresponds to the critical shear stress limit of $\tau_{\rm{cr}} = 0.96 \ \rm{Pa}$ \citep{Skorov2012}.}\label{Reynolds_plot}
\end{figure}

\noindent The relationship of $Re$ to planetesimal size and eccentricity at an orbital distance of $r=0.2 \ \rm{au}$ is shown in Fig. \ref{Reynolds_plot}. The black curve corresponds to the limit set by the critical wall shear stress ($\tau_{\rm{cr}} = 0.96 \ \rm{Pa}$), below which gas erosion does not occur. Thus, Fig. \ref{Reynolds_plot} shows that the relevant Reynolds numbers for the erosion of planetesimals by the gas flow at $r = 0.2 \ \rm{au}$ are $Re \sim 10^{6} - 10^{11}$.

The time for the flow to pass the planetesimal is 

\begin{equation}
t_{\rm{flow}} \sim 3 \times 10^{-1} \ \rm{s} \ \Bigg(\frac{\mathit{e}}{0.1}\Bigg)^{-1} \Bigg(\frac{\mathit{L}}{10^3 \ \rm{m}} \Bigg) \Bigg(\frac{\mathit{M}}{\mathit{M}_{\odot}}\Bigg)^{-1/2} \Bigg(\frac{\mathit{r}}{\rm{au}} \Bigg)^{1/2}.
\end{equation}

\noindent We compare this to the typical rotation timescale of 67P/Churyumov–Gerasimenko of $t_{\rm{rot}}\sim12\ \rm{h}$ \citep{Jorda2016}. Given that $t_{\rm{flow}} < t_{\rm{rot}}$, the flow adapts to the instantaneous direction of the planetesimal. In the case where $t_{\rm{flow}} > t_{\rm{rot}}$, the flow sees the planetesimal shape in an average sense. We note however, that rotation is not considered here.

\subsection{Flow structure}

\begin{figure}[!t]
  \centering
  \includegraphics[trim=60 0 0 0,scale=0.8,clip]{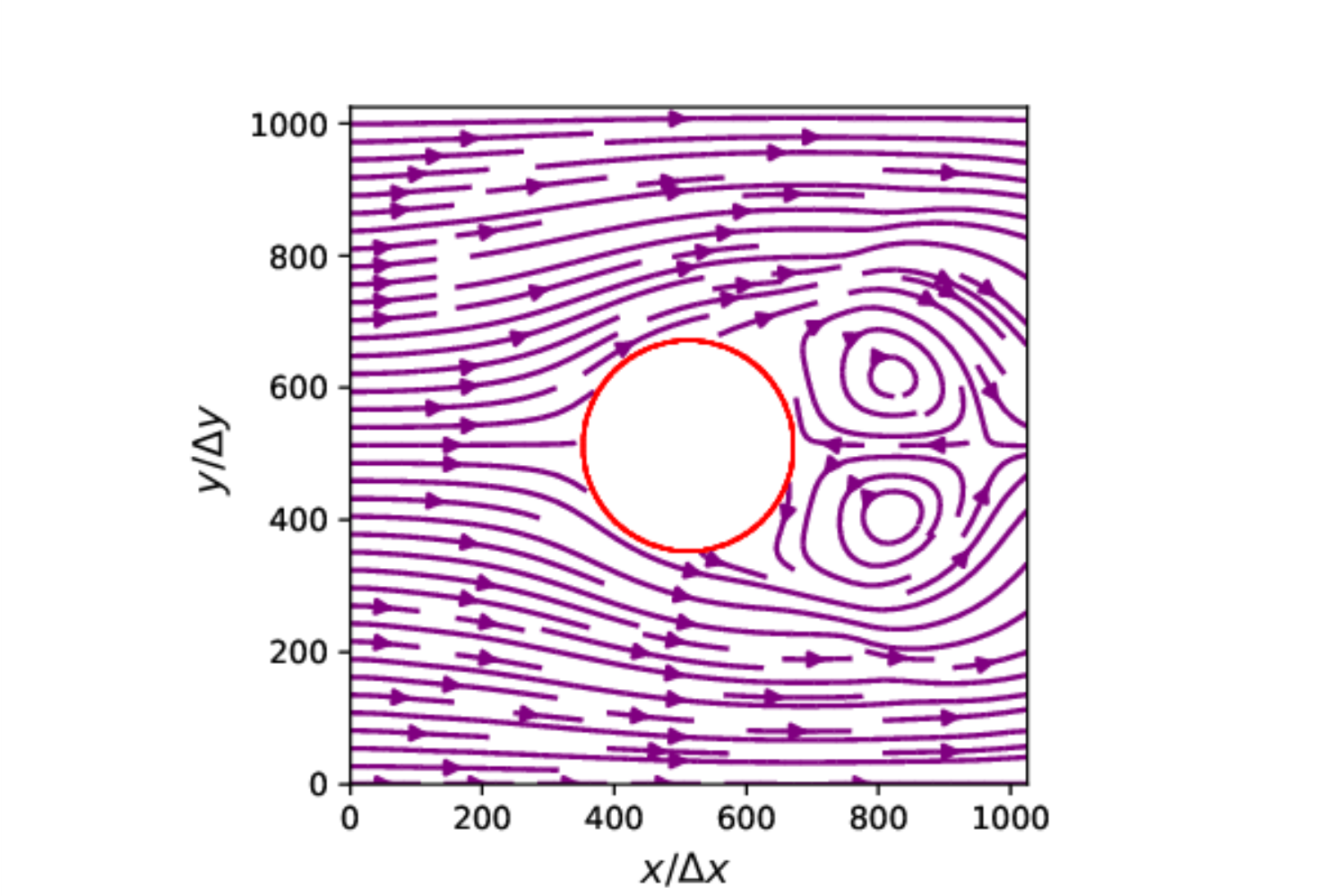}\caption{Streamlines of the flow around a circle (marked with red) given $Re \approx \mathbf{128}$. The streamlines separate in front of the obstacle and two eddies appear behind it.}\label{circle_streamline_Re128}
\end{figure}

The numerical model provides us with a flow profile around the solid obstacle. Given a small Reynolds number, the flow that develops around an obstacle is laminar \citep{Landau1987}. A laminar flow results in smooth, parallel streamlines that separate around the given obstacle. As the Reynolds number increases the laminar flow becomes unstable such as the one presented in Fig. \ref{circle_streamline_Re128}, where $Re \approx \mathbf{128}$ and two eddies form behind the solid. 

\subsection{Wall shear stress profile around solid surface}

\begin{figure}[!t]
  \centering
  \includegraphics[trim=50 0 0 0,scale=0.7,clip]{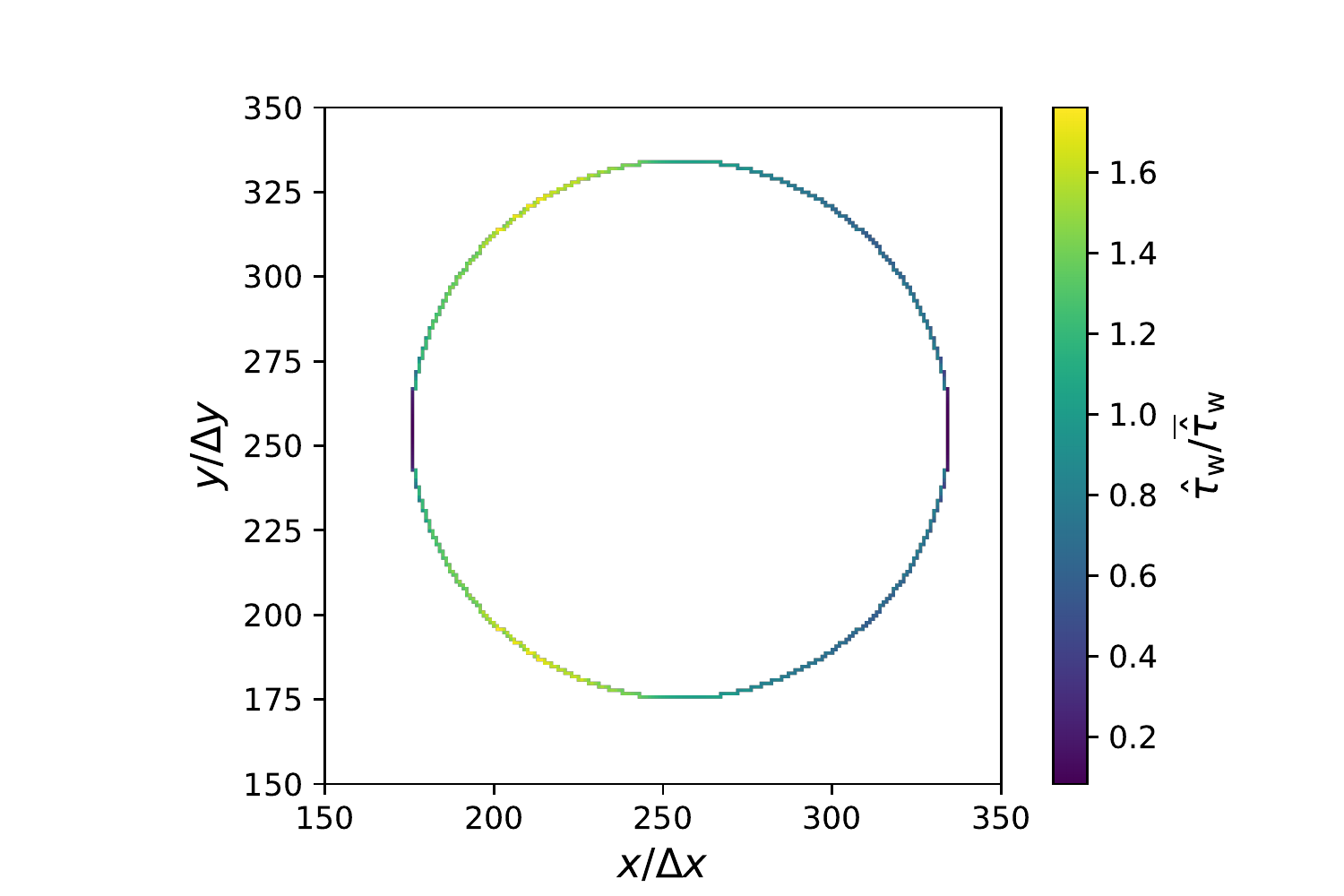}\caption{Wall shear stress over the surface of the obstacle normalized by the mean wall shear stress along the whole surface. The Reynolds number of the flow is $Re \sim 8$. As the wall shear stress is proportial to the mass flux, the erosion is most efficient near the top and bottom of the obstacle (marked with red) and least efficient near the stagnation points of the flow (marked with purple).} \label{circle_mdot}
\end{figure}

We study how the flow affects different parts of the solid by looking at the profile of the wall shear stress along the surface. As the mass flux presented in Eq. \eqref{massloss} is proportional to the wall shear stress, this gives a direct indication of which parts of the obstacle will erode more efficiently. We plot the wall shear stress normalized by the mean stress, $\hat{\tau}_{\rm{w}}/\bar{\hat{\tau}}_{\rm{w}}$, along the surface in Fig. \ref{circle_mdot}. The figure shows that the parts of the solid that are expected to be eroded the least are near the stagnation points of the flow (shown with blue). The wall shear stress is highest (shown with red) near the top of the circle, after which it decreases slightly again at the very top. As expected from the symmetry of the model setup, the wall shear stress profile is symmetric along the horizontal bisector that passes through the midpoint of the circle.

\begin{figure}[!t]
  \centering
  \includegraphics[width=1\columnwidth]{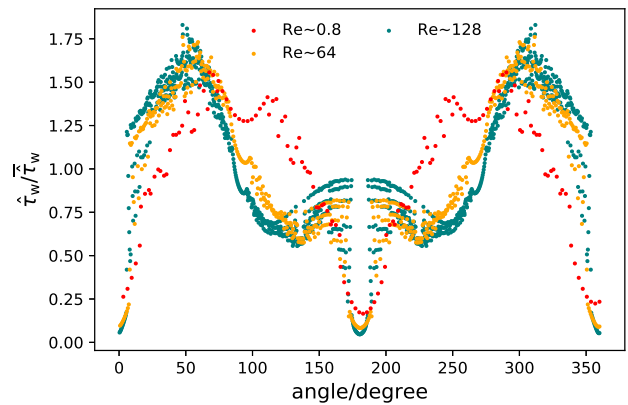}\caption{Wall shear stress along the surface of the obstacle as a function of angle. The angle of $0^{\circ}$ corresponds to the stagnation point on the upstream side of the surface in Fig \ref{circle_mdot}. The different colors correspond to different Reynolds numbers. The pattern of erosion as a result of the wall shear stress exerted by the gas is not a significant function of the $Re$. }\label{circle_mdot_degree}
\end{figure}

To understand how changing the Reynolds number of the flow affects the erosion rate, we compare $\hat{\tau}_{\rm{w}}/\bar{\hat{\tau}}_{\rm{w}}$ for varying values of $Re$ around the surface in Fig. \ref{circle_mdot_degree}. We see that the pattern of the exerted stress changes somewhat as a function of the Reynolds number. The maxima of the curves increase when comparing low Reynolds number flows ($Re < 1$) to the ones with $Re > 1$. The maxima saturate to similar wall shear stress values as instabilities set in ($Re > 1$).

\section{Summary and discussion}
\label{summary}

Our results show that planetesimals can erode by the action of the gas disk in the inner parts of protoplanetary disks, $r$$\lesssim$$1 \ \rm{au}$. As giant planets migrate through the disk and shepherd the planetesimals in front of them, they leave them with excited eccentricities \citep{Tanaka1999}, which facilitates their erosion. We find that gas erosion affects small plantesimals of $\sim$$100 \ \rm{m}$ on orbits with eccentricities larger than $0.1$ and large planetesimals of $\sim$$100 \ \rm{km}$ on orbits with eccentricities above $0.5$. Outside of $r$$\sim$$1 \ \rm{au}$, most disk solids are not largely affected by gas erosion. We also find that if gas erosion is effective, the erosion time of a given planetesimal is short. 

To illustrate the erosion profile along planetesimals, we developed a numerical code using the lattice Boltzmann method (see Appendix \ref{Appendix1}). From the numerical model we found that based on the distribution of the wall shear stress along the solid surface erosion is fastest near the top and bottom of the obstacle and slowest near the stagnation points of the flow. We also find that the Reynolds number of the flow does not significantly alter the pattern of erosion along the surface within the range of Reynolds numbers considered here. Our results lead us to predict that no planetesimal will be found in eccentric orbits in the inner parts of a protoplanetary diks.

This conclusion is subject to some caveats that we discuss below. Firstly, we estimate the wall shear stress via the ratio of the headwind velocity and the thickness of the boundary layer. The latter can have quite complex behaviour. As the planetesimal faces the headwind, there must be a stagnation point at the front. If we assume that the boundary layer remains laminar, its thickness grows as the square root of the distance from the stagnation point measured along the direction of the flow along the planetesimal surface. Let us use the symbol $\xi$ for this coordinate, i.e., $\delta \sim \sqrt{\xi}$. However, this dependence is not true close to the stagnation point. Here, the boundary layer thickness becomes a non-zero constant set by the flow at the stagnation
point \citep{Landau1987}. As $\xi$ increases, we surmise the wall shear stress to increase, reach a maximum value, and then to decrease roughly as $\sqrt{\xi}$, since $\delta$ increases in the same manner. This qualitative behaviour of the wall shear stress is seen in Fig. \ref{circle_mdot}. However, this picture does not hold for large planetesimals or for large Reynolds numbers. In the case of large planetesimals the boundary layer becomes unstable at the critical value of $\xi$. This critical value depends on the mean flow, $U$, and possibly other details, such as the curvature of the planetesimal surface, how smooth or rough the surface is and upon the porosity of the material. For $\xi$ larger than this critical value there is no known simple dependence of the boundary layer thickness on $\xi$. For very large $\xi$ and thus large planetesimals, the boundary layer can be turbulent and the empirical relation $\delta \sim \xi^{1/5}$ may hold. To estimate whether a planetesimal erodes at all we need to determine the maximum wall shear stress. As this estimation is challanging, we choose to use the laminar boundary layer and use the minimum value of the stress thus obtained as our estimate of the stress. Hence, we may grossly underestimate the wall shear stress, particularly for large planetesimals. Thus, it is possible that in reality the planetesimals erode even faster than we estimate here. 

Secondly, as the planetesimal erodes, its shape and size changes. This in turn changes the wall shear stress. We have ignored this effect in our estimates. A more accurate estimate, following \cite{Ristroph2012} will be presented in a future work. However, as we have explained before, for large planetesimals we expect our estimate to provide a lower limit to the wall shear stress. As the size of the planetesimal decreases, our description of the boundary layer as a laminar one becomes more accurate, thus we expect that the inclusion of the dynamical boundary change will make the planetesimal erode also even faster than estimated here. 

Thirdly, we note that for planetesimals on circular orbits, the headwind is in general too small to cause any erosion. However, as $\eta$ (the ratio of the thermal speed of gas molecules
and the Keplerian speed) is very small, even a small eccentricity will give rise to erosion. We expect that planetesimals in the inner disks can survive only if
they are on nearly circular orbits. In the case of a large eccentricity, two additional effects need to be considered, which are not present in our model.
One is that the variation of gas density in the orbit can be quite large. The density is highest where
the speed is largest hence we expect erosion will increase further. The second new effect is that the headwind can become so large that it becomes supersonic. As far as we know, there is no easy way
to estimate the wall stress in such a situation. The Lattice Boltzman method we used in this paper will not work for supersonic flows either. Numerical methods following \citep{Rusanov1976} can be used.
Developing numerical methods for supersonic erosion should be a high priority for future studies of planetesimal erosion.

Finally, the estimate of the threshold stress has a large amount of uncertainty. Our estimates are based on the experiments performed with pebble piles but planetesimals are also expected to have significant amounts of ice in them. The presence of ice may increase the threshold stress manyfold. Furthermore, in a large pile the material inside may become more compact by sintering. Thus the threshold stress for the inner material may be significantly larger than that of the outer layer. Thus, we may find that the erosion is very effective in removing the surface layer of planetesimals but becomes less effective as the inner layers are exposed. We expect our results will encourage further work on this topic

\begin{acknowledgements}

We thank the anonymous referee for their input and comments that helped improve the manuscript. N. S. would like to thank H. Capelo for inspiring discussions. N.S. was funded by the ''Bottlenecks for particle growth in turbulent aerosols'' grant from the Knut and Alice Wallenberg Foundation (2014.0048). A.J. thanks the Swedish Research Council (grant2014-5775), the Knut  and Alice Wallenberg Foundation (grants  2012.0150,2014.0017) and the European Research Council (ERC Consolidator Grant724687-PLANETESYS) for research support.

\end{acknowledgements}

\bibliographystyle{aa}
\bibliography{references}

\begin{appendix}
\section{Lattice Boltzmann method}
\label{Appendix1}
\raggedbottom

The lattice Boltzmann method relies on the kinetic theory \citep{Sukop2007}. The fluid is described as particles that exchange momentum and energy through advection/streaming and collision. These particles are described by the probability distribution function, $f(\mathbf{x,v},t)=f$, that represents the population of particles with position, $\mathbf{x}$ and velocity, $\mathbf{v}$, at time, $t$. The system is then described with the Boltzmann equation, which given that there is no external force acting on the system can be written as

\begin{equation}
\frac{\partial f}{\partial t} + \mathbf{v}\frac{\partial f}{\partial \mathbf{x}} = \varOmega(f),
\label{Boltzmann}
\end{equation}

\noindent where $\varOmega(f)$ is the collision operator. 

\begin{figure}[!t]
\centering
\includegraphics[width=0.8\columnwidth]{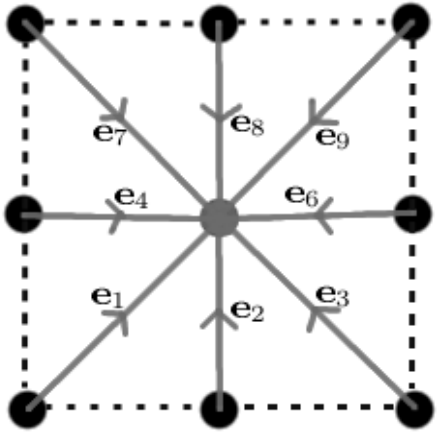}\caption{Illustration of the streaming step in the lattice Boltzmann method. The particle distributions are streamed into the central lattice node from neighbouring nodes with lattice velocities of $\mathbf{e}_a$.}\label{streaming}
\end{figure}

In the lattice Boltzmann method the continuous Boltzmann equation is discretized in both velocity and physical space. The lattice Boltzmann equation is then solved on a lattice, where the particles are restricted to movement along discrete directions between the nearest and next-nearest neighbours of a two-dimensional Cartesian lattice. The particle velocities are thus described by $\mathbf{e}_a$, where $a$ are the direction indexes. As shown in Fig. \ref{streaming}, we use the two dimensional D2Q9 model, such that $a=1,2,...,9$, and $\mathbf{e}_5$ represents a particle at rest (marked as the grey node on Fig. \ref{streaming}).

The advection term in Eq. \eqref{Boltzmann} is implemented as a streaming process, where the probability distributions propagate along the discrete velocity vectors. The collision term in Eq. \eqref{Boltzmann} is most widely approximated with the Bathnagar-Gross-Krook collision operator \citep{Bhatnagar1954}.
Thus, the lattice Boltzmann equation becomes

\begin{equation}
f_a(\mathbf{x}+\mathbf{e}_a \Delta t, t+\Delta{t})=f_a(\mathbf{x},t)-\frac{\Big[f_a(\mathbf{x},t)-f^{\rm{eq}}_a(\mathbf{x},t)\Big]}{\tau},
\label{LBE}
\end{equation}

\noindent where $f_a(\mathbf{x},t)$ describes the directional distribution function. According to Eq. \eqref{LBE}, particles $f_a(\mathbf{x},t)$ move to neighboring lattice nodes at time $t+\Delta{t}$ with velocity $\mathbf{e}_a$, which represents the streaming part of the method. The collision of the particles is described by the second term on the right-hand side of Eq. \eqref{LBE}, where $\tau$ is the relaxation time and  $f^{\rm{eq}}_a(\mathbf{x},t)$ is the equilibrium distribution function given by

\begin{equation}
f^{\rm{eq}}_a(\mathbf{x},t)=w_a \rho \Bigg[1+3\frac{\mathbf{e}_a \cdot \mathbf{u}}{c^2}+\frac{9}{2} \frac{(\mathbf{e}_a \cdot \mathbf{u}^2)}{c^4}-\frac{3}{2}\frac{\mathbf{u}^2}{c^2} \Bigg].
\label{equilibrium}
\end{equation}

\noindent In Eq. \eqref{equilibrium}, $w_a$ represents the weights associated with each direction, such that

\begin{equation}
w_a=
\begin{cases}
4/9, & a=5; \\
1/9, & a=2,4,6,8; \\
1/36, & a=1,3,7,9. \\
\end{cases}
\end{equation}

\noindent In Eq. \eqref{equilibrium}, $\rho=\rho(\mathbf{x},t)$ and $\mathbf{u}=\mathbf{u}(\mathbf{x},t)$ are the macroscopic fluid density and velocity and $c$ is the lattice speed.

The macroscopic fluid density and velocity are calculated using the weighted sums of the probability distribution function's moments, such that the macroscopic fluid density becomes

\begin{equation}
\rho(\mathbf{x},t)= \sum_a f_a(\mathbf{x},t), 
\end{equation}

\noindent and the macroscopic fluid velocity is

\begin{equation}
\mathbf{u}(\mathbf{x},t)=\frac{1}{\rho(\mathbf{x},t)} \sum_a f_a(\mathbf{x},t) \mathbf{e}_a.
\end{equation}

\subsection{Initial and boundary conditions}

The initial conditions of the numerical model are set via the parameters that make up the Reynolds number, which is defined as

\begin{equation}
Re=\frac{u'L'}{\nu'},
\label{Reynolds_code}
\end{equation}

\noindent in terms of the lattice Boltzmann model's parameters. In Eq. \eqref{Reynolds_code} $u'$ is the velocity of the fluid in units of the lattice speed, $c$. In our model the lattice speed is set to $ c=1 $. The planetesimal size, $L'$, is measured in units of the cell size, which we set as $\Delta x = 1$. The lattice viscosity is proportional to the relaxation time, such that

\begin{equation}
\nu'=\frac{1}{3}\Bigg(\tau-\frac{1}{2} \Bigg).
\end{equation}
 
\begin{figure}[!t]
\centering
\includegraphics[trim=50 0 0 15 ,width=1.2\columnwidth,clip]{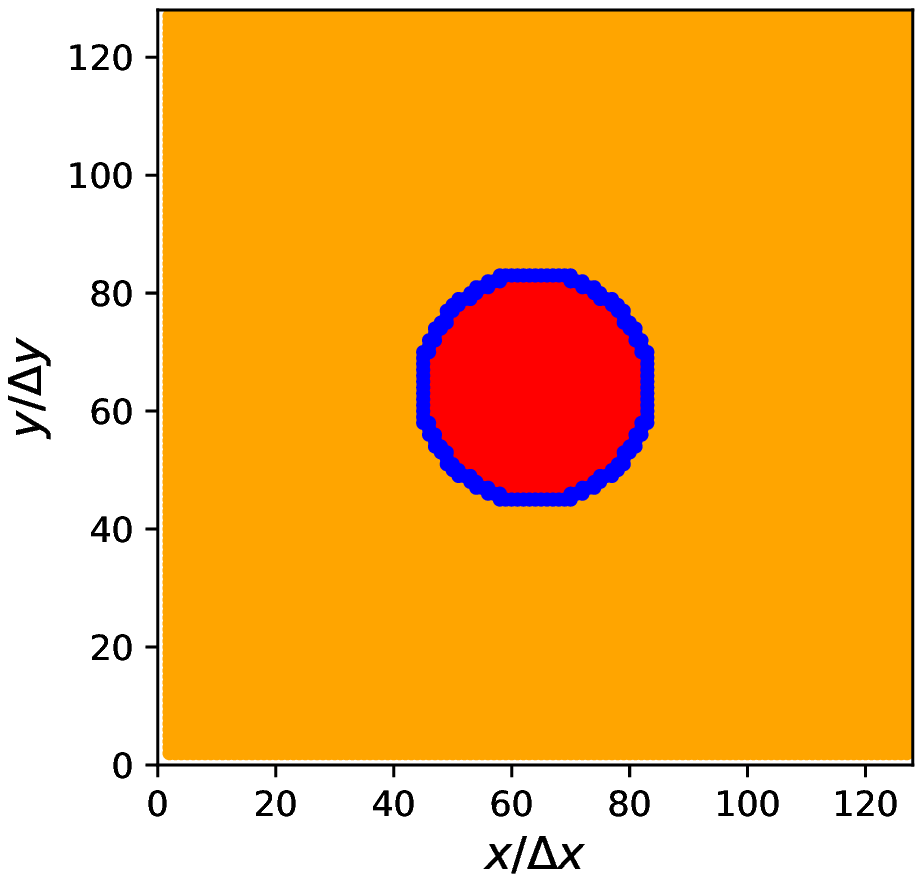}\caption{Example model set-up, where the solid obstacle (red) is surrounded by the gas flow (orange). In between the two lie the surface points (blue), such that the solid points do not direclty come into contact with the flow.}\label{solid_vs_surface}
\end{figure} 

Figure \ref{solid_vs_surface} shows an example set-up, where the solid obstacle is a circle (marked with red) which is surrounded with surface nodes (marked with blue). The surface nodes act as the boundary between the obstacle and the fluid nodes (marked with orange). The lattice Boltzmann equation presented in Eq. \eqref{LBE} is solved on the fluid but not on the solid nodes. The dynamics on the surface nodes are determined by the no-slip boundary condition.

The no-slip boundary condition ensures that particles hitting the surface nodes are bounced back, such that there is no flow into the solid obstacle. The boundary conditions of the model are periodic in the vertical direction such that the distribution functions leaving at the top of the simulation box are the same as the ones at the bottom. Thus, the boundary condition in the vertical direction is such that

\begin{equation}
f_a(x,y_0,t) = f_a(x,y_N,t),
\end{equation}
\begin{equation}
f_a(x,y_{Ny+1},t) = f_a(x,y_1,t),
\end{equation}

\noindent where $y_0$ and $y_{Ny+1}$ are virtual lattice nodes that act as ghost zones around the simulation domain. In the horizontal direction we apply the Dirichlet boundary condition, such that we set the velocity to a constant value at the inlet and the outlet of the simulation domain.  

\subsection{Wall shear stress}

The wall shear stress is calculated using the shear stress tensor defined as 

\begin{equation}
\sigma_{ab} = \mu \Bigg(\frac{\partial u_a}{\partial x_b}+\frac{\partial u_b}{\partial x_a} \Bigg).
\label{sigma_tensor}
\end{equation}

\noindent In the lattice Boltzmann method this is calculated as

\begin{equation}
\sigma_{ab} (\mathbf{x}) = \Bigg(1-\frac{1}{2 \tau} \Bigg) \sum_i f^{\rm{neq}}_{i}(\mathbf{c_i})^a (\mathbf{c_i})^b,
\end{equation}

\noindent where $f^{\rm{neq}}_{i}=f_{i}-f^{\rm{eq}}_{i}$ is the non-equilibrium part of the distribution function \citep{Jager2017}. The force felt by the surface is then the shear tensor multiplied by the surface normal vector, $\hat{n}$, such that

\begin{equation}
\mathbf{F}=\hat{n} \cdot \sigma.
\end{equation}

\noindent Finally, the wall shear stress is the magnitude of the force such that 

\begin{equation}
\tau_{\rm{w}} = | \mathbf{F} |.
\end{equation}

\end{appendix}

\end{document}